\documentclass[10pt, aps, pra, superscriptaddress, 
               nofootinbib, twocolumn, letterpaper]{revtex4-2}
\usepackage{amsmath, amssymb, braket, upgreek, bm}
\usepackage[inline]{enumitem}
\usepackage[dvipsnames]{xcolor}
\usepackage{newtxtext}
\usepackage{booktabs, multirow}
\usepackage{needspace}
\usepackage[unicode=true, bookmarks=true, 
            bookmarksnumbered=false, bookmarksopen=true,
            bookmarksopenlevel=1, breaklinks=false,
            pdfborder={0 0 1}, backref=false,
            colorlinks=true]{hyperref}
\hypersetup{pdfborderstyle=, linkcolor=blue, 
            urlcolor=blue, citecolor=blue}

\usepackage{orcidlink}

\newcommand{\PPO}{\mathrm{C_3 H_6 O}}
\newcommand{\chunk}{\mathrm{C}_3 \mathrm{H}_6^+}
\newcommand{\Oion}{\mathrm{O}^+}
\newcommand{\Cion}{\mathrm{C}^{3+}}

\makeatletter 
\newcommand\semilarge{\@setfontsize\semilarge{11}{13.3}}
\makeatother

\begin{document}

%

\title{Direct imaging of enantiomer-specific orientation dynamics
       \\ in unidirectionally rotating chiral molecules}

\author{Kenta Mizuse\,\orcidlink{0000-0001-8878-808X}}
\email{mizuse@kitasato-u.ac.jp}
\affiliation{Department of Chemistry, School of Science, Institute of Science Tokyo,
             2-12-1-W4-9, Ookayama, Meguro-ku, Tokyo 152-8550, Japan \looseness=-1}
\affiliation{Department of Chemistry, School of Science, Kitasato University,
             1-15-1 Kitazato, Minami-ku, Sagamihara 252-0373, Japan \looseness=-1}

\author{Ilia Tutunnikov\,\orcidlink{0000-0002-8291-7335}}
\affiliation{AMOS and Department of Chemical and Biological Physics,
             The Weizmann Institute of Science, Rehovot 7610001, Israel \looseness=-1}

\author{Long Xu\,\orcidlink{0000-0001-5314-2799}}
\affiliation{Department of Physics, Xiamen University, Xiamen 361005, China}

\author{Yuhei Oyagi}
\affiliation{Department of Chemistry, School of Science, Institute of Science Tokyo,
             2-12-1-W4-9, Ookayama, Meguro-ku, Tokyo 152-8550, Japan \looseness=-1}

\author{Naoya Sakamoto}
\affiliation{Department of Chemistry, School of Science, Institute of Science Tokyo,
             2-12-1-W4-9, Ookayama, Meguro-ku, Tokyo 152-8550, Japan \looseness=-1}
             
\author{Ryo\;Kondo}
\affiliation{Department of Chemistry, School of Science, Institute of Science Tokyo,
             2-12-1-W4-9, Ookayama, Meguro-ku, Tokyo 152-8550, Japan \looseness=-1}

\author{Allan Huang}
\affiliation{Department of Chemistry, The University of British Columbia, 
             Vancouver V6T-1Z1, Canada}

\author{Roman V. Krems}
\email{rkrems@chem.ubc.ca}
\affiliation{Department of Chemistry, The University of British Columbia,
             Vancouver V6T-1Z1, Canada}
\affiliation{Stewart Blusson Quantum Matter Institute, Vancouver, B.C. V6T 1Z4, Canada}

\author{Ilya Sh.\,Averbukh}
\email{ilya.averbukh@weizmann.ac.il}
\affiliation{AMOS and Department of Chemical and Biological Physics,
             The Weizmann Institute of Science, Rehovot 7610001, Israel \looseness=-1}

\affiliation{Department of Chemistry, The University of British Columbia,
             Vancouver V6T-1Z1, Canada}

\author{Yasuhiro Ohshima\,\orcidlink{0000-0002-4826-9004}}
\email{ohshima@chem.titech.ac.jp}
\affiliation{Department of Chemistry, School of Science, Institute of Science Tokyo,
             2-12-1-W4-9, Ookayama, Meguro-ku, Tokyo 152-8550, Japan \looseness=-1}

\date{\today}
\begin{abstract}
Selectively controlling the dynamics of molecular enantiomers underlies advances across chemistry, biology, and physics, yet direct imaging of enantiomer-specific motion has so far remained elusive. Here, we image ultrafast enantioselective orientation dynamics in isolated chiral molecules. Unidirectional coherent rotation induced by a femtosecond laser-pulse pair generates equal and opposite out-of-plane orientations of the two enantiomers. Applying this scheme to 2-methyloxirane, we follow the rotational wave packets by time-resolved Coulomb explosion imaging with two orthogonally arranged detectors. The measured angular distributions reveal that the unidirectional rotation is identical for both enantiomers, while the out-of-plane orientations are mirror images that persist through both early-time quasi-classical and quantum dynamics regimes, in quantitative agreement with simulations. We demonstrate that full angular distributions provide richer dynamical information, with some qualitatively different distributions yielding similar orientation factors upon integration. Our approach opens a route to real-time observation and control of chiral dynamics in the gas phase.
\end{abstract}
\maketitle

\begin{figure}[h!]
    \centering
    \includegraphics[]{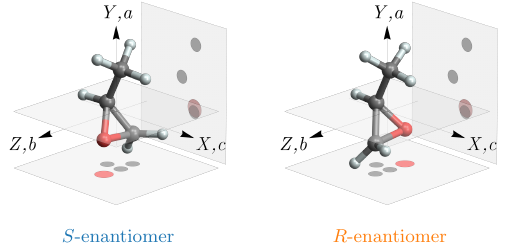}
    \caption{Illustration of the two enantiomers of the chiral 2-methyloxirane molecule 
    $\PPO$ (also known as propylene oxide).
    The $X$, $Y$, and $Z$ axes define the laboratory frame of reference, while $a,\,b,\,c$
    are the molecular principal axes of inertia that define the rotating body-fixed
    frame.
    The atoms are color-coded: red$|$oxygen, grey$|$carbon, off-white$|$hydrogen.
    While the molecular frame (i.e., the $abc$-axes) orientation of the two enantiomers
    is identical, they are \textit{not superimposable}. Red and grey circles
    on the lower $XZ$ plane and the back $XY$ plane are the projections of the
    corresponding spheres (oxygen and carbon atoms, respectively) onto each
    plane. The two enantiomers have \textit{different} ``molecular shadows''
    on the $XZ$ plane but \textit{identical} shadows on the $XY$ plane.
    }
    \label{fig:MOX}
\end{figure}

\noindent
An object that cannot be superimposed on its mirror image is chiral; the two
mirror-image forms are termed left- and right-handed enantiomers. Chirality
is present at all scales, from the spiral arms of galaxies to the helices of DNA and
elementary particles~\cite{Wagniere2007}, and plays a central role in chemistry,
biology, and physics---governing biological processes~\cite{Blackmond2010, Ma2023},
pharmacological effects~\cite{Senkuttuvan2024}, 
catalysis~\cite{Noyori2002, Fanourakis2020}, and certain symmetry violations
in fundamental interactions~\cite{Safronova2018, Quack2022}. Understanding and
exploiting molecular chirality requires an unambiguous specification of, and control
over, molecular handedness. Many observables can be sensitive to
chirality~\cite{Ayuso2022}, but a particularly direct approach to distinguish molecular handedness can be
to fix the spatial orientation of molecules in the $XYZ$ laboratory frame and
attempt to superimpose the enantiomers. As illustrated in Fig.~\ref{fig:MOX}, handedness can be observed in the
``molecular shadows'' that the oriented enantiomers cast onto a chosen plane~\cite{Milner2019, Saribal2021}.

In practice, chirality can be probed with various techniques.
Circular dichroism (CD) spectroscopy, which measures the differential absorption
of left- and right-circularly polarized light, relies on weak mixing of
electric and magnetic transition moments and consequently requires macroscopic
samples with long optical paths or high concentrations. More sensitive approaches
exploit higher-order or coherent processes: photoelectron circular dichroism
(PECD)~\cite{Bowering2001, Lux2012, Janssen2014, Rozen2019, Sparling2025} and
circular dichroism in high-harmonic generation~\cite{Cireasa2015, Baykusheva2018}
encode chirality in the dynamics of ejected or recolliding electrons; Coulomb
explosion imaging (CEI) directly resolves molecular geometry via coincidence
measurements on fragmented ions~\cite{Pitzer2013, Herwig2013}; and microwave
three-wave mixing (MW3WM)~\cite{Hirota2012, Patterson2013a, Patterson2013b, Patterson2014}
exploits dipole-allowed rotational transitions in the microwave domain to
generate phase-sensitive enantioselective polarization along a chosen laboratory
axis. However, with the exception of coincidence-CEI, these techniques compress
the underlying enantiosensitive dynamics into a small number of integrated
observables and provide little direct access to the time-dependent angular
distributions of the rotating molecules.

Theoretical work has predicted that nonresonant laser fields and/or
THz pulses with twisted polarization can drive gas-phase chiral molecules into
enantioselective \textit{molecular orientation}, in which the two enantiomers
point in opposite directions in the laboratory
frame~\cite{Yachmenev2016, Gershnabel2018, Tutunnikov2018, Tutunnikov2021THz}.
An intuitive classical picture of the underlying torque is given in
Refs.~\cite{Gershnabel2018, Tutunnikov2018}. The prediction was confirmed
experimentally using an optical centrifuge~\cite{Milner2019}, available in only
a handful of laboratories. A centrifuge traps the rotational motion of molecules in an accelerating field-induced potential, gradually increasing the rotational period over multiple tens of picoseconds. This timescale, however, precludes access to the subpicosecond
domain, where the enantioselective torque is predicted to be imprinted most directly. An alternative, and far simpler, approach can be based on a twisted-polarization field produced by
a pair of time-delayed, polarization-crossed, linearly polarized pulses. In
non-chiral molecules, such pulse pairs have been shown to excite rotational
wave packets (WPs) exhibiting unidirectional rotation
(UDR)~\cite{Fleischer2009, Kitano2009, Mizuse2015, Lin2015}. UDR has also been
achieved with an optical centrifuge~\cite{Ivanov1999, Villeneuve2000, Yuan2011, Korobenko2014}
and with polarization-shaped pulses~\cite{Karras2015}. However, the feasibility of the latter technique to drive
enantioselective dynamics in chiral molecules has not yet been
experimentally demonstrated.

Here, we report direct imaging of UDR-driven enantioselective orientation
dynamics in a chiral molecule. We apply a pair of time-delayed,
polarization-crossed femtosecond pulses (defining the $XY$ plane in
Fig.~\ref{fig:MOX}) to enantiopure \textit{S}- and \textit{R}-2-methyloxirane
(MOX), and probe the resulting rotational wave packets with time-resolved CEI with
two orthogonally arranged detectors---one resolving the in-plane angular
distribution of $\Cion$ fragments, and the other the out-of-plane distribution of
$\Oion$. We report three observations: first, the unidirectional rotation in the
polarization plane is identical for the two enantiomers, consistent with
the in-plane torque being dominated by polarizability tensor elements
that are invariant under the mirror operation relating \textit{S}- and
\textit{R}-MOX. Second, the out-of-plane
orientations are mirror images that persist through both
early-time, quasi-classical as well as quantum dynamics regimes, in quantitative agreement with numerical simulations of the
time-dependent Schr\"odinger equation. Third, while the orientation
factor $\braket{\cos(\theta)}$ is shown to provide a compact enantioselective scalar observable, the
full angular distributions are illustrated to reveal additional structure, including delays at which similar orientation factors mask
qualitatively different distributions. The approach uses standard tools of femtosecond
optics, applies, in principle, to any gas-phase chiral molecules, and opens a novel
route to real-time observation and control of enantiomer-specific dynamics.\\

\begin{figure}[h]
    \centering
    \includegraphics{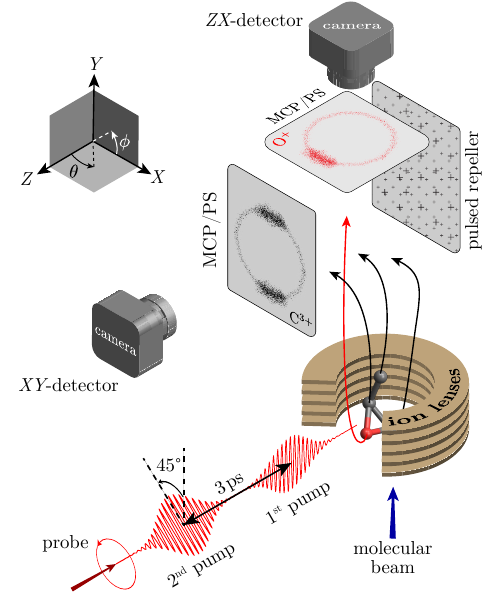}
    \caption{Schematics of the experimental setup, defining the laboratory $XYZ$
    reference frame. The system includes two sets of multichannel plate/phosphor screens
    (MCP/PS) with a camera, enabling the independent and simultaneous imaging of $\Cion$ and
    $\Oion$ ions. All laser pulses propagate against the $Z$ axis. The polarization of the first pump pulse is parallel to the $Y$ axis, while the polarization of the second
    delayed pulse is tilted by $45^\circ$ from the $Y$ axis in the $XY$ plane. 
    This pair of pulses induces unidirectional molecular rotation in the $XY$ plane 
    and orientation along/against the $Z$ axis. 
    The CEI probe pulse is circularly polarized in the $XY$ plane.
    }
    \label{fig:exp_setup}
\end{figure}

\noindent {\large \textbf{Results}}\\[0.5em]
\noindent {\semilarge \textbf{Mirror-symmetric unidirectional rotation}}\\*[0.5em]
\noindent Figure~\ref{fig:exp_setup} shows the experimental setup, including the dual-angle
ion imaging apparatus and the laboratory coordinate system used in this study. 
Here, unidirectional rotation is induced in the $XY$ plane, with enantioselective dynamics
expected along the $Z$ axis.
\begin{figure*}[ht!]
    \centering
    \includegraphics{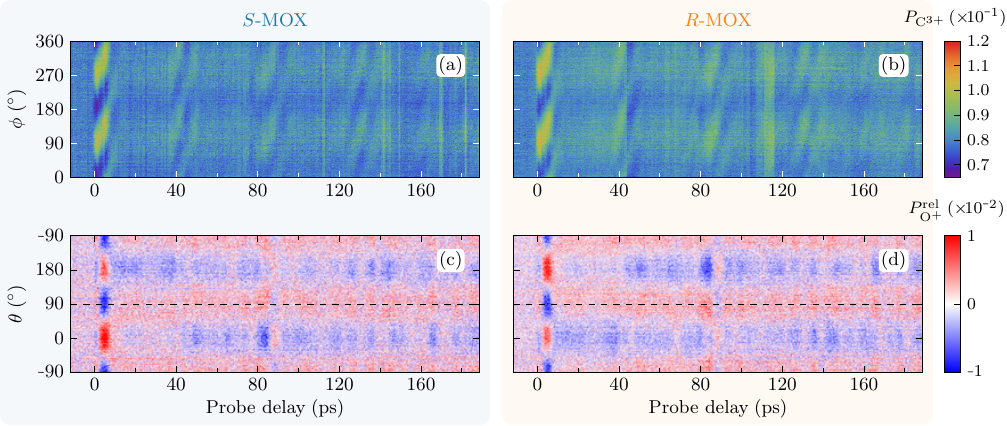} 
    \caption{Observed time-dependent angular distributions. The first pump inducing alignment
    was applied at the $\text{delay} = 0$, and the second pump inducing the directional control
    was applied at $\text{delay} = 3.0$\,ps. Angular distribution of $\Cion$ in the $XY$ plane,
    $P_{\Cion}(\phi,t)$ is shown in panels (a) \textit{S}-MOX and (b) \textit{R}-MOX. 
    Relative angular distribution of $\Oion$ in the $XZ$ plane, $P^{\mathrm{rel}}_{\Oion}(\theta,t)$ 
    is shown in panels (c) \textit{S}-MOX and (d) \textit{R}-MOX.
    }
    \label{fig:exp-main-result}
\end{figure*}
We employed a previously reported pump-probe optical setup and a molecular beam source
~\cite{Mizuse2015, Mizuse2017}, while the dual-angle ion imaging system was newly installed.
Details of the experiments are described in Methods.
In short, an adiabatically cooled sample of enantiomerically pure \textit{S}- or 
\textit{R}-MOX entrained in He buffer gas was introduced as a molecular beam, directed
towards the $+Y$ axis, and irradiated successively by three femtosecond laser pulses.
The first one, linearly polarized along the $Y$ axis ($\phi = 90^\circ$) was used for
molecular alignment. The second, a delayed replica of the first one but with linear polarization
tilted $45^\circ$ from the $Y$ axis ($\phi = 135^\circ$), was for initiating the UDR-WP dynamics.
The third, circularly polarized in the $XY$ plane, served as the CEI probe.
Upon the probe irradiation, the sample molecules were multiply ionized and exploded into
various fragment ions within the laser duration, the so-called Coulomb explosion (CE).
The ions were accelerated towards the $+Y$ direction and were spatially separated based
on their mass-to-charge ratio and initial velocity. In the present study, $\Cion$ ions
and  $\Oion$ ions were monitored, respectively, to track the UDR and the time-dependent
orientational dynamics of the enantiomers. The former ions were repelled to the $+Z$
direction by applying a pulsed high voltage to the repeller and detected by the $XY$
detector. The $XZ$ detector captured the latter ions. Both the $XY$ and $XZ$
images obtained were 2D projections of portions of 3D Newton spheres, and were used to
evaluate the 2D distributions within the planes.

Figures~\ref{fig:exp-main-result}(a) and  \ref{fig:exp-main-result}(b) show false-color
2D representation (``quantum carpet'') of the distribution of $\Cion$ in the $XY$ plane, 
$P_{\Cion}(\phi,t)$, as a function of probe delay, measured for \textit{S}- and 
\textit{R}-MOXs, respectively. The distribution of the $\Cion$ ions was taken as a measure
of the UDR of the $\mathrm{C\text{--}C\text{--}C}$ backbone and the molecular $a$ axis.
After the first pump pulse, the angular distribution began to exhibit anisotropy.
The alignment parameter 
$\braket{\cos^2 (\phi)}_{\Cion} \equiv \int \cos^2 (\phi) P_{\Cion}(\phi,t)\,d\phi$
reached its maximum ($\approx 0.54$) during the first $1\text{--}3\,\mathrm{ps}$.
After the second polarization-tilted pulse at a delay of $3\,\mathrm{ps}$, 
the light and dark blue areas shift towards larger values of $\phi$, 
resulting in the diagonal stripes visible in Figs.~\ref{fig:exp-main-result}(a) and 
\ref{fig:exp-main-result}(b). 
Such a spatiotemporal pattern in the quantum carpet for the angular distribution is 
a signature of the UDR-WP dynamics, as previously seen in linear molecules
~\cite{Mizuse2015, Lin2015, Xu2020}. 
\begin{figure}[h]
    \centering
    \includegraphics{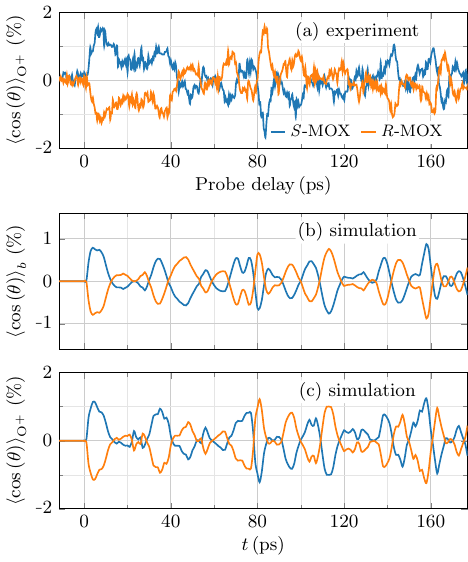} 
    \caption{Time variation of the orientation factors for \textit{S}- and \textit{R}-MOXs:
    (a) observed values for $\braket{\cos (\theta)}_{\Oion}$,
    (b) simulated (by solving TDSE) $\braket{\cos (\theta)}_b$, where $\theta_b$
    is the angle between the $Z$ axis and the projection of the $b$ axis onto the $XZ$ plane,
    (c) calculated $\braket{\cos (\theta)}_{\Oion}$ from the CE trajectories of $\Oion$ ions.
    \vspace{-5mm}
    }
    \label{fig:orientation}
\end{figure}

After $\approx 7\,\mathrm{ps}$, the contrast of the stripes degrades, reflecting the dispersion 
of the WP. In the case of the UDR-WP of linear molecules, a dispersed WP was localized
again after evolving for a specific time, i.e., the \textit{rotational revival period}
~\cite{Seideman2005, Ohshima2010, KochLemeshko2019}. 
Then, the WP exhibits unidirectional rotation while repeatedly expressing localized and
dispersed characters of the angular distribution. In the present case, since the rotational constants of MOX 
(inversely proportional to the corresponding moments of inertia) are
in the relation $A \gg B \approx C$ (Methods, 
Table~\ref{tab:inertia_polarizability}), the molecule is close to a prolate symmetric top,
and has an approximate rotational revival at delay of $1/(B+C) \approx 80\,\mathrm{ps}$. 
During the delays equal to a fraction of the approximate revival time,
($\text{delay} \approx 45,\,85,\,125$, and $165\,\mathrm{ps}$), the stripes' contrast
is partially recovered, indicating some localization, but the degree of recovery
degrades over time. The revivals' imperfections, in contrast to linear molecules, are 
a characteristic feature for the rotational dynamics of an asymmetric top molecule, 
whose rotational energy levels are not in harmonic relation~\cite{ZareBook}, preventing
the rotational WPs from fully returning to the initial state~\cite{Felker1992, Poulsen2004, Seideman2005}.
Typical chiral molecules are asymmetric tops. The unidirectional rotation of the
chiral molecule is clearly visible in Figs.~\ref{fig:exp-main-result}(a) and
\ref{fig:exp-main-result}(b), and no detectable difference between the two
enantiomers is present in the in-plane dynamics. This is consistent with the
in-plane torque being dominated by polarizability tensor elements (notably the
diagonal $\alpha_{aa}$; see Methods, Table~\ref{tab:inertia_polarizability})
that are invariant under the mirror operation relating \textit{S}- and
\textit{R}-MOX, while the parity-flipping off-diagonal elements
($\alpha_{ab}, \alpha_{bc}$) contribute only as a small correction
in the polarization plane. Enantiomer-specific behavior must therefore be
sought in the out-of-plane motion, where these off-diagonal elements drive
opposite torques on the two enantiomers (Discussion).\\

\noindent {\semilarge \textbf{Enantiomer-dependent dynamics}}
\\[0.5em]
\noindent Figures~\ref{fig:exp-main-result}(c) and \ref{fig:exp-main-result}(d) show the angular
distributions of $\Oion$ ions in the $XZ$ plane for \textit{S}- and \textit{R}-MOX, respectively.
Since the change in the angular distribution from an isotropic state was quite small, we present the
\textit{relative} distribution, 
$P^{\mathrm{rel}}_{\Oion}(\theta,t) \equiv P_{\Oion}(\theta,t)-\int \! P_{\Oion}(\theta,t)\,d\theta$,
where $P_{\Oion}(\theta,t)$ stands for the angular distribution of $\Oion$, and the subtracted second term is the isotropic offset (the angle-integrated probability density).  
As seen clearly, $P^{\mathrm{rel}}_{\Oion}(\theta,t)$ remains isotropic after the interaction with
the first pulse, but shows a significant change following the second pump pulse. 
The distribution becomes concentrated around $\theta = 0^\circ,\,180^\circ$
(red spots), parallel or antiparallel to the laser propagation direction ($\pm Z$). 
This concentration reaches a maximum at the time when the UDR-WP begins to disperse 
(around $7\,\mathrm{ps}$), as shown in Figs.~\ref{fig:exp-main-result}(a) and \ref{fig:exp-main-result}(b).
Then, the red area quickly shifts toward $\theta=90^\circ,\,270^\circ$
and then persists, exhibiting some modulation.
This means that the distribution for the ejected direction of $\Oion$ stays concentrated near the $XY$ plane. 
The distribution again focuses along the $Z$ axis 
(i.e., $\theta = 0^\circ,\,180^\circ$) at $\approx 87\,\mathrm{ps}$, which is close
to the first full revival after the initial maximum orientation (at $\approx 7\,\mathrm{ps}$).
Similar to the alignment of $\Cion$, the revival is imperfect, and the 
contrast of $P^{\mathrm{rel}}_{\Oion}(\theta,t)$ at $\approx 87\,\mathrm{ps}$ is significantly
lower compared to $7\,\mathrm{ps}$.

We next focus on asymmetry in the quantum carpets above and below the horizontal line at 
$\theta = 90^\circ$, corresponding to the forward and backward orientation along the $Z$ axis. 
A close inspection of Figs.~\ref{fig:exp-main-result}(c) and \ref{fig:exp-main-result}(d) 
reveals a small but clear difference around $\theta=0^\circ,\,180^\circ$ for
the two enantiomers at early times ($\text{delay} \approx 3\text{--}7\,\mathrm{ps}$). 
For the \textit{S} enantiomer, the probability of $\Oion$ being close to 
$\theta=0^\circ$ is slightly larger (i.e., redder and wider) than 
$\theta=180^\circ$.
For the \textit{R} enantiomer,
the probability distribution shows the \textit{opposite trend}.  
During the probe delay interval of $10\text{--}45\,\mathrm{ps}$, the area within 
$\theta=180^\circ \pm 45^\circ$ for \textit{S}-MOX is bluer (i.e., showing lesser probabilities)
than that within $\theta= 0^\circ \pm 45^\circ$, while the opposite is seen in \textit{R}-MOX. 
At the half revival ($\approx 45\,\mathrm{ps}$), the bluer areas switch from lower to upper
or vice versa. Near the full revival ($\approx 87\,\mathrm{ps}$), the probability distribution
around $\theta=0^\circ$ becomes minimal (bluest) for \textit{S}-MOX, whereas that around 
$\theta=180^\circ$ becomes minimal for \textit{R}-MOX. To summarize, anti-correlated behaviors
for the \textit{S} and \textit{R} enantiomers were clearly identified in their time-dependent
angular probability distributions probed as $\Oion$ images in the $XZ$-plane.

For a quantitative discussion of the enantioselective orientational dynamics, the orientation
parameter, $\braket{\cos(\theta)}_{\Oion}\equiv \int \! \cos(\theta)P^{\mathrm{rel}}_{\Oion}(\theta,t)\,d\theta$, is evaluated. 
Figure~\ref{fig:orientation}(a) shows the time-dependent orientation factor 
$\braket{\cos(\theta)}_{\Oion}$ for the two enantiomers of MOX, extracted from the observed
angular distribution of $\Oion$ [Figs.~\ref{fig:exp-main-result}(c) and \ref{fig:exp-main-result}(d)]. 
The parameter remains zero until unidirectional rotation begins after the second pump
pulse is introduced. After the second pulse, $\braket{\cos{(\theta)}}_{\Oion}$ for the
two enantiomers exhibits pronounced oscillations, changing sign over the entire observed
time window.
Most importantly, the signals from the two enantiomers are \emph{perfect} 
reflections to each other across the horizontal line $\braket{\cos(\theta)}_{\Oion} = 0$.\\

\noindent{\semilarge \textbf{Quantum simulations}}
\\[0.5em]
\noindent While the early-time orientation following the onset of UDR is captured
by a classical model (Discussion below), a full quantum treatment is required to
reproduce the structured time evolution observed thereafter. We therefore solve
the time-dependent Schr\"odinger equation (TDSE) for a rigid asymmetric-top MOX
molecule subject to the two pump pulses; the asymmetric-top parameters and
numerical procedure are detailed in Methods.

\begin{figure}[h!]
    \centering
    \includegraphics[]{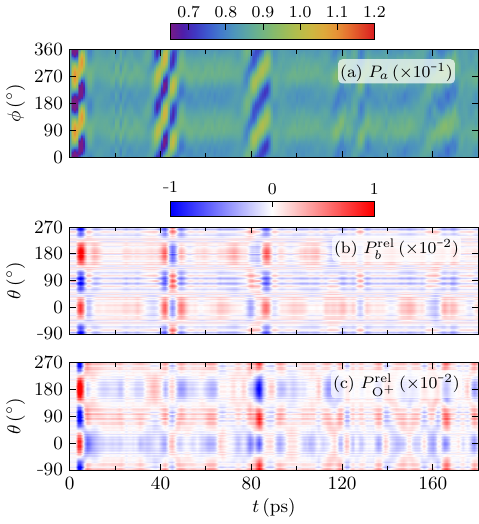}
    \caption{Simulation results for $R$-MOX. (a) probability density $P_a(\phi,t)$ for the projection of the molecular $a$ axis onto the $XY$ plane. The probability that the projection lies between $\phi$ and $\phi +d\phi$ is $P_a(\phi,t)\,d\phi$.
    (b) relative probability density $P^{\mathrm{rel}}_b(\theta,t)$ for the projection of the molecular $b$ axis onto the $XZ$ plane.
    (c) relative probability distribution $P^{\mathrm{rel}}_{\Oion}(\theta,t)$,
    obtained from the CE trajectories simulations.
    }
    \label{fig:simulation}
\end{figure}

Figure~\ref{fig:simulation}(a) shows the simulated probability density
$P_a(\phi,t)$ for the projection of the molecular $a$ axis onto the $XY$ plane,
computed for \textit{R}-MOX. Because the in-plane unidirectional rotation is
enantiomer-independent (as mentioned above), the corresponding distribution for
\textit{S}-MOX is identical, and the simulation reproduces the diagonal-stripe
quantum-carpet pattern of Figs.~\ref{fig:exp-main-result}(a) and
\ref{fig:exp-main-result}(b) for both enantiomers. This confirms that the
$\Cion$ ejection direction faithfully tracks the $a$-axis orientation.

The relative probability density $P^{\mathrm{rel}}_b(\theta,t)$ for the projection of
the $b$ axis onto the $XZ$ plane is shown in Fig.~\ref{fig:simulation}(b),
again for \textit{R}-MOX. The calculated distribution for \textit{R}-MOX is the
mirror image of that for \textit{S}-MOX across $\theta = 90^\circ$, so the
\textit{R}-simulation should match the observed \textit{R}-MOX angular
distribution [Fig.~\ref{fig:exp-main-result}(d)], with the (unshown)
\textit{S}-simulation matching Fig.~\ref{fig:exp-main-result}(c) by reflection.
The simulated and observed distributions are in close correspondence, consistent
with the O atom in MOX lying approximately along the $b$ axis
(Fig.~\ref{fig:MOX}). The simulated orientation factor $\braket{\cos(\theta)}_b$,
plotted in Fig.~\ref{fig:orientation}(b), reproduces the measured
$\braket{\cos(\theta)}_{\Oion}$ [Fig.~\ref{fig:orientation}(a)] across both
quasiclassical and revival regimes.

To complete the link between molecular orientation and the imaged $\Oion$
distribution, we additionally simulate classical CE trajectories for the
$\Oion$ fragment; the procedure and assumptions are detailed in Methods. The
resulting relative probability distribution $P^{\mathrm{rel}}_{\Oion}(\theta,t)$
[Fig.~\ref{fig:simulation}(c)] and the corresponding orientation factor
[Fig.~\ref{fig:orientation}(c)] agree with the observed distributions still
better than the bare $b$-axis projection, verifying that the ejected direction
of $\Oion$ is a quantitative proxy for the underlying enantioselective MOX
orientation.\\

\noindent {\large \textbf{Discussion}}\\[0.5em]
\noindent{\semilarge \textbf{Classical mechanism of enantioselective torque}}
\\*[0.5em]
\noindent To give an intuitive picture of the early-time dynamics we follow the
analysis of laser-induced enantioselective orientation in twisted-polarization
fields developed in Refs.~\cite{Gershnabel2018, Tutunnikov2018}. In
implementations based on a pair of strong, time-delayed
pulses~\cite{Yachmenev2016, Gershnabel2018, Tutunnikov2018}, the orientation
develops on short enough timescales that a classical description is justified
in the early-time regime.
\begin{figure*}[t]
    \includegraphics{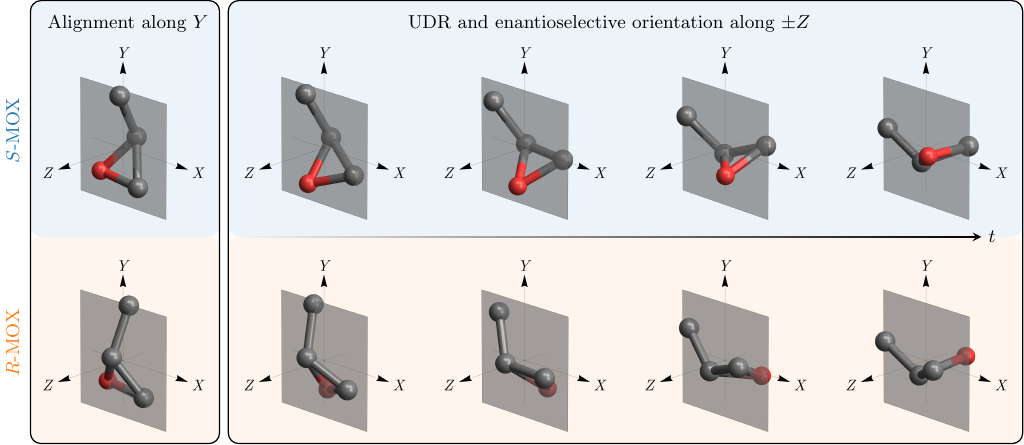}
    \caption{
    Qualitative illustration of the orientation mechanism.
    The two rows follow two sample MOX molecules (\textit{S} and \textit{R} enantiomers).
    Hydrogens are hidden to avoid clutter.
    The first image in each row shows an idealized scenario where the most polarizable
    molecular axis (approximately the principal axis of inertia, $a$) is perfectly aligned
    along the laboratory $Y$ axis, i.e., the polarization direction of the first pulse.
    Note: alignment means that for each molecule with $a$ axis along $+Y$, there is a
    partner molecule oriented along $-Y$ (not shown).
    The sequential images along the time arrow show the dynamics following the second pulse,
    which induces unidirectional molecular rotation in the $XY$ plane.
    The additional enantioselective torque applied around the $a$ axis of the chiral
    molecules forces them to twist out of the $XY$ plane, such that the oxygen atom turns
    towards $+Z$ for the \textit{S} enantiomer, and towards $-Z$ for the \textit{R} enantiomer.
    }
\label{fig:theo-mechanism}
\end{figure*}
In our experiments, the first femtosecond pulse, polarized along $Y$, polarizes
the molecules through their polarizability tensor
$\hat{\boldsymbol{\alpha}}$. The interaction between the induced dipole
$\mathbf{d}_{\rm ind} = \hat{\boldsymbol{\alpha}}\mathbf{E}$ and the field
$\mathbf{E}$ produces a torque equivalent to an impulsive kick on the most
polarizable molecular axis towards $\pm Y$. Since this axis is close to the
principal axis of inertia $a$ (Fig.~\ref{fig:MOX}), the $a$ axis aligns along $Y$
shortly after the kick (first column of Fig.~\ref{fig:theo-mechanism};
see~\cite{Seideman2005, LemeshkoKrems2013, KochLemeshko2019} for reviews of
laser alignment). Alignment by a single pulse is not enantioselective.

The second pulse, applied at the moment of maximal alignment along $Y$ and
polarized at $45^\circ$ between $Y$ and $-X$, induces UDR of the most
polarizable axis about $Z$. As shown in Fig.~\ref{fig:theo-mechanism}, this
in-plane motion is identical for both enantiomers, as previously observed in
non-chiral
molecules~\cite{Fleischer2009, Kitano2009, Korech2013, Mizuse2015, Lin2015}. An
additional torque, however, is specific to chiral molecules: it depends on the
off-diagonal elements of the polarizability tensor, which carry opposite signs
for the two enantiomers (Methods, Table~\ref{tab:inertia_polarizability}). This
makes the torque enantioselective, driving the oxygen atom toward $+Z$ for the
\textit{S} enantiomer and toward $-Z$ for the \textit{R} enantiomer
(Fig.~\ref{fig:theo-mechanism}). The predicted sign matches the experimental
sign of $\braket{\cos(\theta)}_{\Oion}$ at early times
[Fig.~\ref{fig:orientation}(a)].\\

\noindent{\semilarge \textbf{Quantum dynamics and revivals}}
\\*[0.5em]
\noindent The classical picture above captures the early-time enantioselective
torque but cannot account for the structured time evolution observed at longer
delays. Two distinct quantum-coherence regimes are visible in the data and
reproduced by the TDSE simulations. The first is the dispersion of the
rotational wave packet beginning at $\approx 7\,\mathrm{ps}$, after which the
orientation factor $\braket{\cos(\theta)}_{\Oion}$ remains nonzero and oscillates
in time [Fig.~\ref{fig:orientation}(a,b)] while the in-plane distributions lose
their diagonal-stripe contrast. The second is the partial revival near
$\approx 87\,\mathrm{ps}$, set by the asymmetric-top rotational period
$1/(B+C)$. Because MOX is an asymmetric top whose rotational levels are not in
harmonic relation~\cite{ZareBook}, full revivals of the rotational wave packet
are precluded~\cite{Felker1992, Poulsen2004, Seideman2005}; the imperfect
revivals seen in Figs.~\ref{fig:exp-main-result} and \ref{fig:simulation} are
characteristic of asymmetric-top dynamics. That the simulated and measured
distributions agree across both regimes establishes that dual-detector CEI
captures the rotational wave-packet evolution from the quasiclassical kick
through the deep-quantum revival.\\

\noindent{\semilarge \textbf{Limitations of scalar orientation factor}}
\\*[0.5em]
\noindent Previous studies of enantioselective orientation have largely relied
on the orientation factor $\braket{\cos(\theta)}$ as the primary
observable~\cite{Milner2019, Gershnabel2018, Tutunnikov2018, Tutunnikov2021THz}.
This compresses the underlying dynamics into a single signed scalar and, as our
data show, conceals important structure. While $\braket{\cos(\theta)}_{\Oion}$
remains positive (negative) for \textit{S}-MOX (\textit{R}-MOX) over
$5\text{--}45\,\mathrm{ps}$, the underlying angular distributions in
Figs.~\ref{fig:exp-main-result}(c) and \ref{fig:exp-main-result}(d) exhibit
qualitative changes near $\approx 10\,\mathrm{ps}$ that the scalar metric
cannot register.

\begin{figure*}[ht!]
\centering
\includegraphics{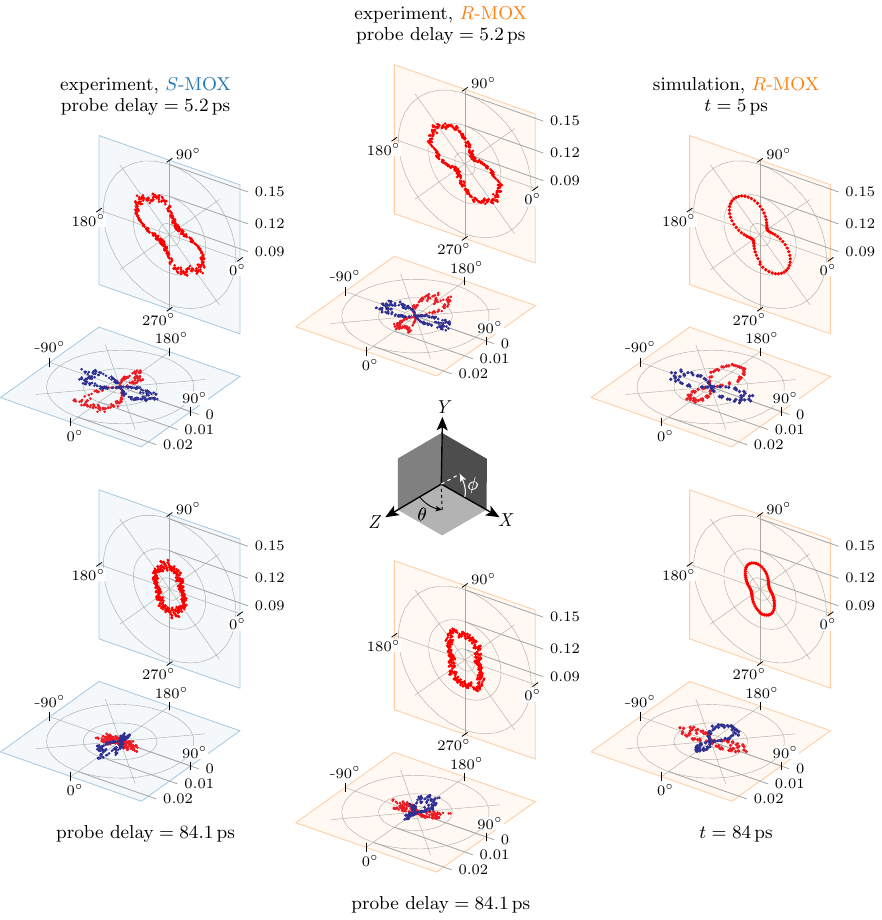}
\caption{
    Polar plots for the observed $P_{\Cion}(\phi)$ in the $XY$ plane and
    $P^{\mathrm{rel}}_{\Oion}(\theta)$ in the $XZ$ plane for \textit{S}-MOX (left),
    \textit{R}-MOX (middle), and the corresponding simulated ones for \textit{R}-MOX (right)
    at $t = 5.2\,\mathrm{ps}$ (upper) and $t = 84.1\,\mathrm{ps}$ (lower), respectively.
    Note that scales for $P_{\Cion}(\phi)$ are in the range of
    0.08 to 0.15, while $P^{\mathrm{rel}}_{\Oion}(\theta)$ is indicated as its absolute
    value spanned within 0\textendash 0.02, with red and blue dots corresponding to positive
    and negative values.
    }
\label{fig:2D_images}
\end{figure*}

A particularly clean illustration is provided by the two delays at which the
orientation factor reaches its largest magnitudes, $t = 5.2\,\mathrm{ps}$ and
$t = 84.1\,\mathrm{ps}$. At both times
$|\braket{\cos(\theta)}_{\Oion}| \approx 0.015$, yet the angular distributions
are qualitatively different (Fig.~\ref{fig:2D_images}): at
$5.2\,\mathrm{ps}$ the factor is dominated by a relatively positive
distribution along $Z$, whereas at $84.1\,\mathrm{ps}$ it arises from a
relatively negative distribution along $Z$. The same scalar can correspond to
physically distinct angular landscapes. Direct mapping of the full angular
distribution---rather than reduction to a single moment---is therefore
essential for resolving the dynamics, especially in the deep-quantum regime
where multiple modes contribute coherently.\\

\noindent{\semilarge \textbf{Limitations and outlook}}
\\*[0.5em]
\noindent Several aspects of the present experiments limit the information that
can be extracted. First, after adiabatic cooling the rotational distribution
retains a finite temperature of a few Kelvins, which broadens the wave packet
and reduces the contrast of revivals~\cite{Felker1992, Seideman2005}; lower
temperatures, e.g., from a buffer-gas-cooled source or  rotational-stale selection, e.g., by Stark decelerator \cite{Meerakker2008} or deflector \cite{Filsinger2009}, would sharpen both the
quantum-carpet stripes and the orientation factor. Second, the assignment of
$\Cion$ to the $a$ axis and $\Oion$ to the $b$ axis is approximate, and the
non-coincidence imaging used here cannot, on its own, distinguish between
fragmentation channels with the same mass-to-charge ratio. A coincidence-CEI
implementation would relax these assumptions and provide direct access to the
full 3D molecular orientation rather than 2D projections. Third, the strong
breaking of cylindrical symmetry inherent to the twisted-polarization scheme
prevents 3D reconstruction from a single 2D image; orthogonal-detector imaging
recovers the additional information needed for the present analysis but does
not yield the complete 3D angular distribution.

These limitations point to natural extensions. The same pulse-pair scheme
should apply to any gas-phase chiral molecule whose polarizability anisotropy
is sufficient to drive UDR; replacing the second pulse with a tailored
polarization shape would allow targeted manipulation of specific quantum
revivals. Most importantly, the high temporal resolution and direct angular
access demonstrated here open a route from \textit{observation} to
\textit{control}---real-time, enantioselective shaping of rotational wave
packets using only standard tools of femtosecond optics, with implications for
gas-phase chiral chemistry, attosecond probing of chiral electron dynamics,
and ultimately for chirality-resolved molecular separation.\\

\noindent {\large \textbf{Materials and Methods}}\\[0.5em]
\noindent{\semilarge \textbf{Experimental design}}\\*[0.5em]
\noindent The objective of the experiments was to image the time-dependent
angular distributions of methyloxirane molecules subjected to a pair of
polarization-crossed femtosecond pulses, and to compare the in-plane ($XY$) and
out-of-plane ($XZ$) distributions for the two enantiomers. Measurements on
enantiopure \textit{S}- and \textit{R}-MOX samples were carried out under
identical conditions. Angular distributions are direct observables and no
statistical hypothesis testing is employed; for each probe delay we accumulated
a fixed number of laser shots and normalized the resulting 2D images to those
obtained under probe-only conditions to remove detector inhomogeneity. Because
the same circularly polarized probe was applied to both enantiomers, any
residual circular dichroism in the CE process cancels in the per-enantiomer
normalization.\\

\noindent{\semilarge \textbf{Experimental procedure}}
\\*[0.5em]
\noindent Trace amounts ($\approx 0.3\%$) of enantiomerically pure \textit{S}-
or \textit{R}-MOX vapor (purity $>98\%$), entrained in He buffer gas at
$3\,\mathrm{MPa}$, were introduced into the vacuum chamber through an
Even--Lavie pulsed valve operated at $250\,\mathrm{Hz}$. The molecular beam,
directed along the $+Y$ axis, passed through two 2.0-mm-diameter skimmers and a
buffer chamber before entering the imaging region; the rotational temperature
of the adiabatically cooled molecules was estimated to be a few Kelvins. The
target molecules were then irradiated by three femtosecond pulses from a
Ti:sapphire amplifier (Quantronix Odin-II HE; $2\,\mathrm{mJ}$ per pulse,
$500\,\mathrm{Hz}$).

The first (alignment) pulse, linearly polarized along the $Y$ axis
($\phi = 90^\circ$), had a center wavelength of $810\,\mathrm{nm}$, a duration
of $300\,\mathrm{fs}$, and a peak intensity below $10\,\mathrm{TW/cm^2}$. The
second pulse, a delayed replica of the first but with linear polarization
tilted $+45^\circ$ from the $Y$ axis ($\phi = 135^\circ$), initiated the
UDR-WP dynamics. The pulse separation was set to $3.0\,\mathrm{ps}$, the time
of maximum alignment of the C--C--C backbone along $Y$ as determined
experimentally from the time-dependent $\Cion$ distribution after the first
pulse alone. The third (probe) pulse, circularly polarized in the $XY$ plane
($\sim 60\,\mathrm{fs}$, $\sim 800\,\mathrm{TW/cm^2}$), drove Coulomb explosion.

The dual-angle imaging apparatus uses two 2D position-sensitive detectors with
orthogonal imaging planes. Each consists of a chevron microchannel-plate stack
(MCP, $75\,\mathrm{mm}$ active diameter) backed by a phosphor screen and read
out by a lens-coupled digital camera. The $XZ$ detector is perpendicular to the
ion-lens axis (a standard velocity-map-imaging geometry). The $XY$ detector
lies parallel to both the ion-lens axis and the $XY$ plane, offset
$\sim 40\,\mathrm{mm}$ from the central axis to avoid intercepting ions
destined for the $XZ$ detector; a pulsed repeller electrode placed
$80\,\mathrm{mm}$ in front of the $XY$ detector pushes selected ions onto its
face when activated.

Upon probe irradiation the sample molecules were multiply ionized and exploded
into various fragment ions, which were accelerated to $\sim 3\,\mathrm{keV}$
along $+Y$ and separated by mass-to-charge ratio. Initially all detector
components were at ground potential. When $\Cion$ ions reached the front of
the $XY$ detector, high-voltage pulses were applied to the pulsed repeller
($+5500\,\mathrm{V}$), the MCP ($-720\,\mathrm{V}$), and the phosphor
($-3200\,\mathrm{V}$) of the $XY$ detector, pushing $\Cion$ onto its surface;
the pulses (typical duration $\sim 300\,\mathrm{ns}$) were turned off again
before slower ions arrived. We selected $\Oion$ rather than O$^{2+}$ to
minimize $XY$-imaging-induced perturbations of the ion trajectories. After
the $XY$ imaging, $\Oion$ ions traversed the $XY$ detector in the field-free
state and were detected on the $XZ$ detector, which was operated with a
$\sim 70\,\mathrm{ns}$ gate pulse to discriminate $\Oion$ from neighboring
$m/z$ species (notably CH$_3^+$).

Both $XY$ and $XZ$ images are 2D projections of portions of 3D Newton spheres.
For the $XY$ image, the circularly polarized probe acts as a window function:
the anisotropic ionization probability favors fragments ejected within the
polarization plane, so the $XY$ images are effectively slices through the $XY$
plane ($\theta = 90^\circ$). For the $XZ$ image, we estimate that
$\sim 50\%$ of the ion cloud contributes, given the time-slice scheme. Because
the present dynamics lack cylindrical symmetry, full 3D reconstruction from a
single image is not possible; the recorded 2D image is used directly to
extract the $\theta_{\Oion}$ distribution. Detector inhomogeneities were
compensated by normalizing all angular distributions to those obtained under
probe-only conditions; this also cancels any residual circular dichroism in
the CE process, since the same circularly polarized probe was applied to both
enantiomers.\\

\noindent{\semilarge \textbf{Quantum simulation of rotational dynamics}}
\\[0.5em]
\noindent To describe the laser-driven rotational dynamics of MOX we use the
rigid asymmetric-top model. The Hamiltonian is given by~\cite{ZareBook}
\begin{equation}
   \hat{\mathcal{H}}_R
   =
   \sum_{q = a,b,c}\frac{\hat{J}_q ^2}{2 I_q},
\end{equation}
where $\hat{J}_q$ is the angular-momentum operator along the principal axis of
inertia $q$ and $I_q$ is the corresponding moment of inertia. For a generic
asymmetric top $I_a < I_b < I_c$. MOX is close to a prolate top because
$I_b \approx I_c$, making it convenient to write the matrix representation of
$\hat{\mathcal{H}}_R$ ($\mathbf{H}_R$) in the prolate symmetric-top basis
$\ket{JKM}$, where $J$ is the total angular-momentum quantum number and $K$,
$M$ are its projections on the principal axis $a$ and the laboratory axis $Z$,
respectively.

The energy levels of the asymmetric top and the corresponding eigenstates are
computed numerically by diagonalizing $\mathbf{H}_R$. The eigenvector elements
of $\mathbf{H}_R$ are the coefficients $c_K^{J,\tau,M}$ used to construct the
rotational states of MOX as
$\ket{J\tau M} = \sum_K c_K^{J,\tau,M}\ket{JKM}$.

The light--matter interaction is described in the length gauge as
\begin{equation} \label{eq:H-int}
    \hat{\mathcal{H}}_{\rm int}(t)
   =
   -\hat{\boldsymbol{\upmu}}\cdot \mathbf{E}(t)
   -
   \frac{1}{2} [\hat{\boldsymbol{\alpha}} \mathbf{E}(t)] \cdot \mathbf{E}(t),
\end{equation}
where $\hat{\boldsymbol{\upmu}}$ is the molecular dipole operator,
$\hat{\boldsymbol{\alpha}}$ the molecular polarizability tensor operator, and
$\mathbf{E}(t)$ the laser electric field. Our experiments use femtosecond
pulses at optical frequency $\omega$ (far-detuned from the rotational
transitions) such that $\mathbf{E}(t)\propto\cos(\omega t)$, so we average the
Hamiltonian over the optical cycle to obtain
$\hat{\mathcal{H}}_{\rm int}(t) =
-[\hat{\boldsymbol{\alpha}} \bm{\mathcal{E}}(t)] \cdot \bm{\mathcal{E}}(t)/4$,
where $\bm{\mathcal{E}}(t) = \mathbf{E}(t)/\cos(\omega t)$ is the envelope
function. The dipole term $-\hat{\boldsymbol{\upmu}}\cdot \mathbf{E}(t)$
vanishes after averaging.

When the eigenstates of the full Hamiltonian
$\hat{\mathcal{H}} = \hat{\mathcal{H}}_{R} + \hat{\mathcal{H}}_{\rm int}(t)$
are expanded in the $\ket{JKM}$ basis, the TDSE takes the form of a system of
coupled differential equations. During the pulses we solve the coupled
equations numerically. During the field-free stages we expand the state
vectors in the asymmetric-top eigenstates (eigenvectors of $\mathbf{H}_R$) and
propagate the nonstationary states analytically. Observables of interest are
computed for each initial state $\ket{J\tau M}$ and the final
finite-temperature result is obtained by Boltzmann averaging. For the pulse
intensities used here and the estimated rotational temperature of
$4\,\mathrm{K}$, the basis includes all $\ket{JKM}$ with $J \leq 12$, and the
thermal average runs over initial states $\ket{J\tau M}$ with $J \leq 8$.

Table~\ref{tab:inertia_polarizability} lists the molecular parameters of
\textit{R}-MOX used in the simulations. \textit{S}-MOX is obtained from
\textit{R}-MOX by a reflection in a mirror plane; assuming the mirror is set
in the $ac$ plane (see Fig.~\ref{fig:MOX}), the off-diagonal elements
$\alpha_{ab}$ and $\alpha_{bc}$ of \textit{S}-MOX have opposite signs. The
choice of mirror plane does not affect physical observables. The simulation
parameters used: peak intensity and FWHM duration of both pulses are
$3\,\mathrm{TW/cm^2}$ and $300\,\mathrm{fs}$, with a $3\,\mathrm{ps}$ delay
between them.

\begin{table}[h]
\centering
\caption{Values of the molecular moments of inertia and polarizability tensor elements
for $R$-MOX (in atomic units).
Rotational constants ($A$, $B$, and $C$) are inversely proportional
to the corresponding moments of inertia.}
\label{tab:inertia_polarizability}

\setlength{\tabcolsep}{8pt}
\renewcommand{\arraystretch}{1.2}
\begin{tabular*}{\linewidth}{@{\extracolsep{\fill}}lll@{}}
\toprule
\multirow{2}{*}{Moments of inertia} &
\multicolumn{2}{c}{Polarizability tensor elements} \\
\cmidrule(l){2-3}
& Diagonal & Off--diagonal \\
\midrule
$I_a = 180\,386$ & $\alpha_{aa}=45.63$ & $\alpha_{ab}=2.56$ \\
$I_b = 493\,185$ & $\alpha_{bb}=37.96$ & $\alpha_{ac}=0.85$ \\
$I_c = 553\,513$ & $\alpha_{cc}=37.87$ & $\alpha_{bc}=0.65$ \\
\bottomrule
\end{tabular*}
\end{table}

\noindent{\semilarge \textbf{Coulomb-explosion trajectory simulation}}
\\[0.5em]
\noindent To convert the simulated MOX orientation distribution into a
predicted $\Oion$ angular distribution, we performed classical CE trajectory
simulations under simplified assumptions. First, we consider only the double
ionization of MOX, the lowest-order (and presumably dominant) ionization
pathway producing $\Oion$ by CE. Second, we assume that following the probe
pulse the MOX molecule dissociates instantaneously into a point-mass $\Oion$
ion and a rigid $\chunk$ fragment containing the remaining atoms; this is
consistent with an independent synchrotron-radiation study showing that the
channel $\PPO + h\nu \rightarrow \chunk + \Oion + 2e$ is one of the dominant
two-body fragmentation channels of MOX~\cite{Falcinelli2018}. Since $\chunk$ is
much heavier than $\Oion$, $\chunk$ is treated as stationary on the relevant
timescale---set by the time for $\Oion$ to reach its asymptotic velocity.

The trajectory of $\Oion$ is then evaluated by placing a unit positive charge
on each atom of $\chunk$ in turn and solving the classical equations of motion
for the asymptotic $\Oion$ velocity. The final velocity vector is computed as
the mean of the nine vectors so obtained. Convolving the result with the
time-dependent probability density for the spatial orientation of MOX
(from the TDSE solutions) yields the relative probability distribution
$P^{\mathrm{rel}}_{\Oion}(\theta,t)$ used for comparison with experiment
[see Figs. \ref{fig:orientation}(c) and \ref{fig:simulation}(c)].

\bibliography{bibliography}
\vspace{0.5em}
\noindent{\semilarge \textbf{Acknowledgments}}
\\*[0.5em]
\noindent The authors thank Mr.\ Hiroki Haneda for his contribution to the
early stage of this study. I.S.A.\ gratefully acknowledges the hospitality
extended to him during his stay at the Department of Chemistry of the
University of British Columbia. This research was made possible in part by
the historic generosity of the Harold Perlman Family.\\

\noindent\textbf{Funding:}
JSPS KAKENHI Grants JP25H01271, JP24K08358, and JP23H03994 (K.M.).
JSPS KAKENHI Grants JP18H03897, JP20K21169, JP22H00312, JP22K18327, and
JP25K22240 (Y.Oh.).
JSPS Core-to-Core Program JPJSCCA20210004 and JPJSCCA20240002 (K.M., Y.Oh.).
Morino Foundation for Molecular Science (K.M.).
JST FOREST Program Grant JPMJFR236P (K.M.).
Natural Sciences and Engineering Research Council of Canada (R.V.K.).\\

\noindent\textbf{Author contributions:} K.M., I.T., R.V.K., I.S.A.,
and Y.Oh. conceived this collaborative experimental-theoretical study.
R.V.K., I.S.A., and Y.Oh. supervised the work. K.M., Y.Oy., N.S., and
R.Ko. conducted the experiments. I.T., L.X., and A.H. performed the
theoretical simulations. K.M. and I.T. prepared the figures. K.M.,
I.T., R.V.K., I.S.A., and Y.Oh. wrote the original draft; reviewed
and edited the manuscript. All authors offered constructive feedback
on the manuscript.\\

\noindent\textbf{Competing interests:} The authors declare no competing
interests.\\

\noindent\textbf{Data and materials availability:} All data needed to evaluate
the conclusions of the paper are present in the paper. The raw experimental and
simulation data underlying all figures in the paper (except the schematic
diagrams) are available in the Zenodo repository at 
\href{https://doi.org/10.5281/zenodo.20829237}{doi:10.5281/zenodo.20829237}.

\end{document}